\documentclass[iop]{emulateapj-rtx4}

\usepackage{graphicx}  
\usepackage{dcolumn}   
\usepackage{bm}        
\usepackage{amsfonts,amsmath,amssymb,mathrsfs}
\usepackage{color}

\usepackage{natbib,times}
\definecolor{orange}{rgb}{1,0.5,0}

\newcommand{\cf}{\textit{cf.}~}
\newcommand{\ie}{\textit{i.e.}~}
\newcommand{\eg}{\textit{e.g.}~}

\newcommand{\hz}{{\rm Hz}}

\shorttitle{Instability-driven evolution of magnetic fields in
  relativistic stars} 
\shortauthors{Ciolfi et al.}

\begin{document}

\title{Instability-driven evolution of poloidal magnetic fields in
  relativistic stars}

\author{Riccardo Ciolfi\altaffilmark{1}, Samuel
  K. Lander\altaffilmark{2,1}, Gian Mario Manca\altaffilmark{1},
  Luciano Rezzolla\altaffilmark{1}}

\email{ciolfir@aei.mpg.de}

\altaffiltext{1}{Max-Planck-Institut f\"ur Gravitationsphysik,
  Albert-Einstein-Institut, Potsdam, Germany}

\altaffiltext{2}{School of Mathematics, University of Southampton,
  Southampton, UK}

\begin{abstract}
The problem of the stability of magnetic fields in stars has a long
history and has been investigated in detail in perturbation
theory. Here we consider the nonlinear evolution of a nonrotating
neutron star with a purely poloidal magnetic field, in general
relativity. We find that an instability develops in the region of the
closed magnetic field lines and over an Alfv\'en timescale, as
predicted by perturbation theory. After the initial unstable growth,
our evolutions show that a toroidal magnetic field component is
generated, which increases until it is locally comparable in strength
with the poloidal one. On longer timescales the system relaxes to a
new non-axisymmetric configuration with a reorganization of the
stellar structure and large-amplitude oscillations, mostly in the
fundamental mode. We discuss the energies involved in the instability
and the impact they may have on the phenomenology of magnetar flares
and on their detectability through gravitational-wave emission.
\end{abstract}
		

\section{Introduction}

During at least two points within a neutron star's (NS) life,
large-scale magnetic field rearrangement may occur. These are shortly
after the formation of NSs in supernovae~\citep{Bonanno:2003uw}, and
also during the giant flares of
magnetars~\citep{ThompsonDuncan1996,Geppert2006}. Whilst similar
rearrangements may also occur in other stars, they are likely to be
particularly significant for the physics of NSs, where the fields are
exceptionally strong: up to $10^{13}\,$G at the surface of normal
pulsars and $10^{15}\,$G for magnetars. There are a variety of
instabilities in NSs, and in a proto-NS magnetic fields may actually
have a stabilising effect (see,
\eg~\citet{Miralles2002,Bonanno:2003uw}), but we are concerned here
with the fast-acting ``Tayler instability'' which affects purely
poloidal (or purely toroidal) magnetic fields in stars.

The magnetic-field geometry of a NS is important for the star's
evolution, provides a distortion that may lead to gravitational
radiation~\citep{Bonazzola1996}, as well as powering the mechanisms by
which these stars may be observed: the pulsar emission for normal NSs,
and the X/$\gamma$-ray emission of magnetars. It is important
therefore to determine which models of magnetised NSs are stable
equilibria.

The study of magnetised stellar equilibria dates back
to~\citet{Chandrasekhar1953}. Since then, many possible magnetic
equilibria have been studied, using both analytic and numerical
techniques. These have included configurations with purely poloidal
fields~\citep{Ferraro1954, Monaghan1965, Bocquet1995} and purely
toroidal fields~\citep{Roxburgh1963, Kiuchi2008}, as well as mixed
poloidal-toroidal configurations~\citep{Roxburgh1966, Haskell2008,
  Tomimura2005, Lander:2009, Ciolfi2009, Ciolfi2010}.

However, constructing a configuration in equilibrium is only half the
problem when modelling stellar magnetic fields; one also needs them to
be stable over many dynamical timescales, since stellar magnetic
fields have been observed to be long-lived. This has proved to be a
challenging problem for analytic methods, which can only study the
initial localised instability and not the resultant field
configuration. With purely poloidal and purely toroidal fields known
to be unstable~\citep{Markey1973, Wright1973, Tayler1973,
  Flowers1977}, only mixed-field configurations are likely to exist in
stars.

More recently it has become feasible to use numerical evolutions to
study these hydromagnetic instabilities, with the benefit that the
global behaviour of the instability may be studied (analytic works
rely on local analyses), as well as the final outcome of the
instability when the field undergoes significant
rearrangement~\citep{Lander:2011, Braithwaite2007, Geppert2006,
  Kiuchi2011}. Despite this recent progress, there are still very few
models of stellar magnetic-field configurations whose stability has
been assessed.

The instability-induced redistribution of magnetic flux is potentially
a very violent event and it has been suggested as a trigger mechanism
for the giant flares of magnetars~\citep{ThompsonDuncan1996}. This
redistribution is likely to be accompanied by a significant change to
the mass quadrupole moment of a NS, making it a potentially detectable
source of gravitational waves (GWs)~\citep{Kashiyama2011,
  Corsi2011}. For this reason it is important to understand the
frequency, amplitude and duration these GWs may have.

This paper is organised as follows. In Section 2 we give a description
of our computational infrastructure and initial stellar models. In
Section 3 we present results from our evolutions, showing the
generation of an instability and the subsequent reorganisation of the
magnetic field into a more stable configuration. We also study the GW
emission from the instability and assess its
detectability. Conclusions are presented in Section 4.

\begin{figure*}
  \begin{center}
     \includegraphics[angle=0,width=5.4cm]{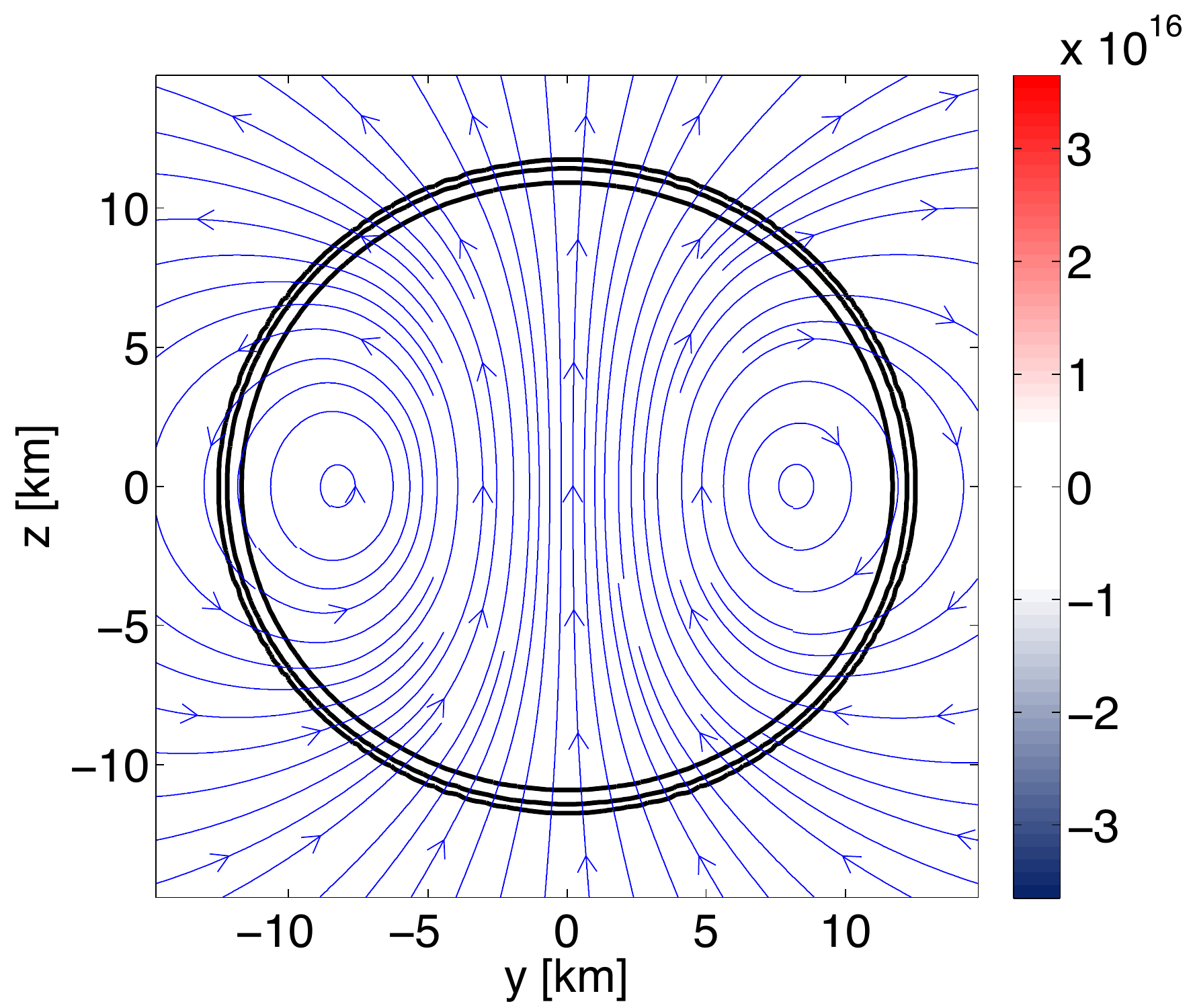}
     \hskip 0.cm
     \includegraphics[angle=0,width=5.4cm]{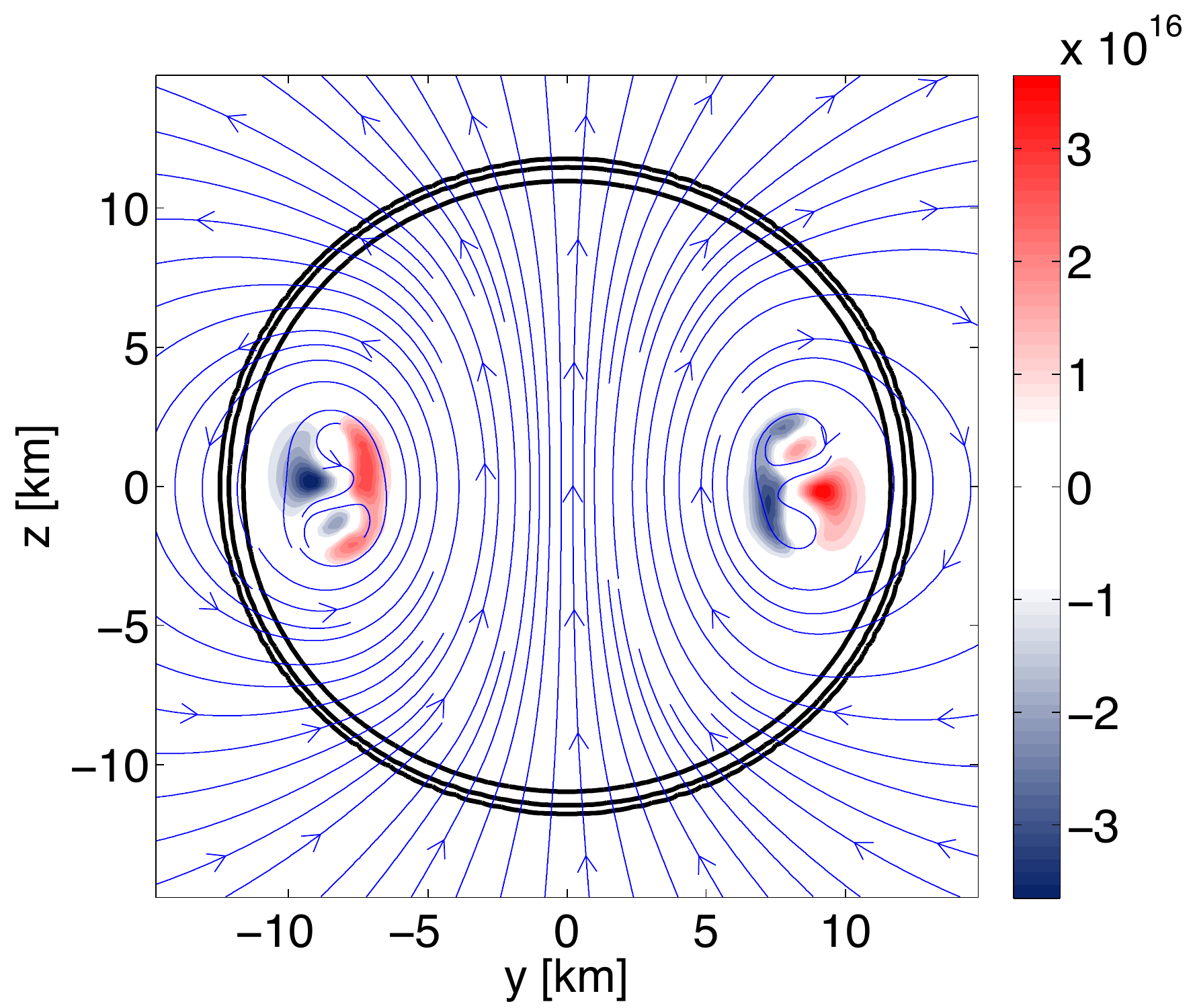}
     \hskip 0.cm
     \includegraphics[angle=0,width=5.4cm]{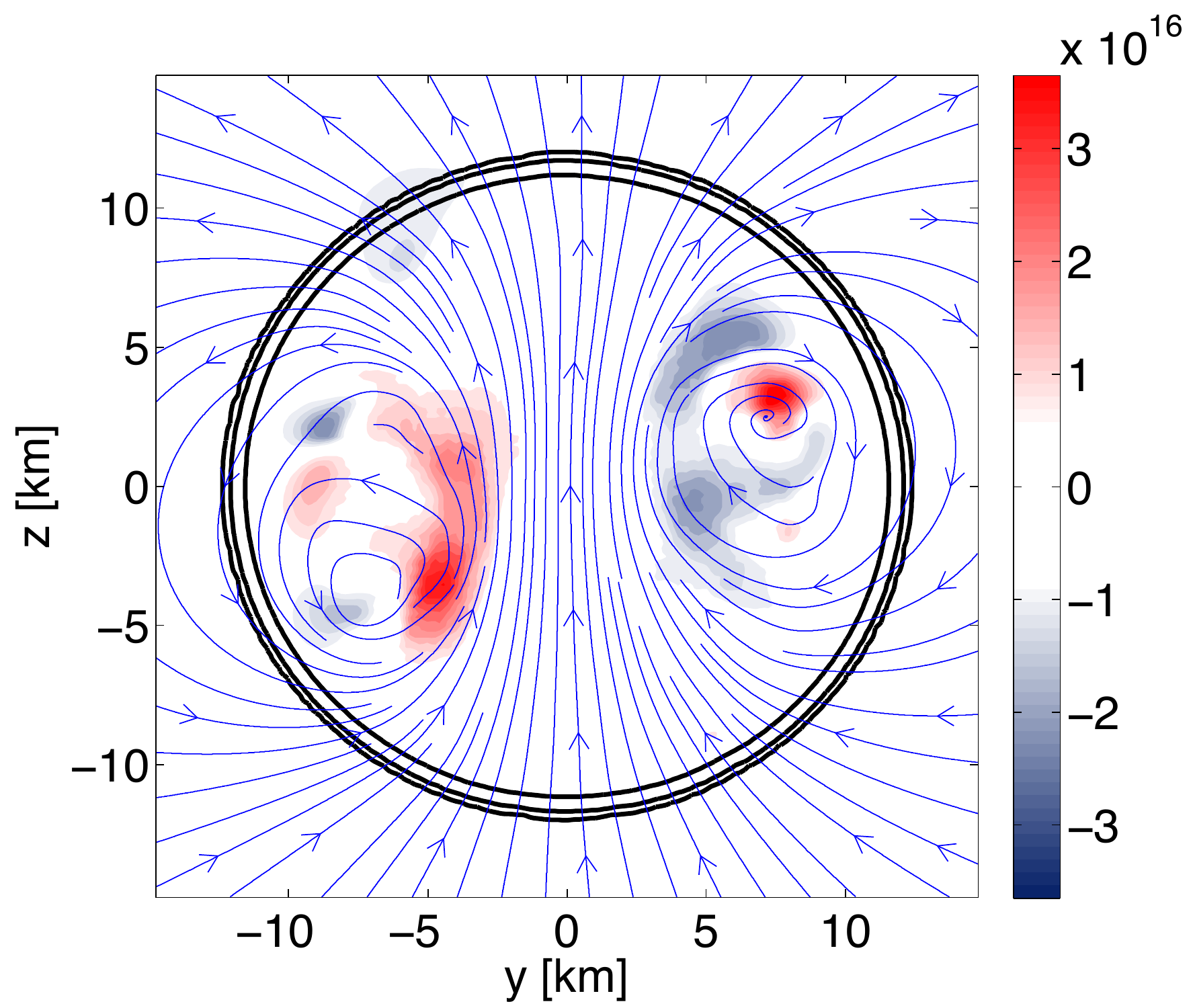}
     \includegraphics[angle=0,width=5.4cm]{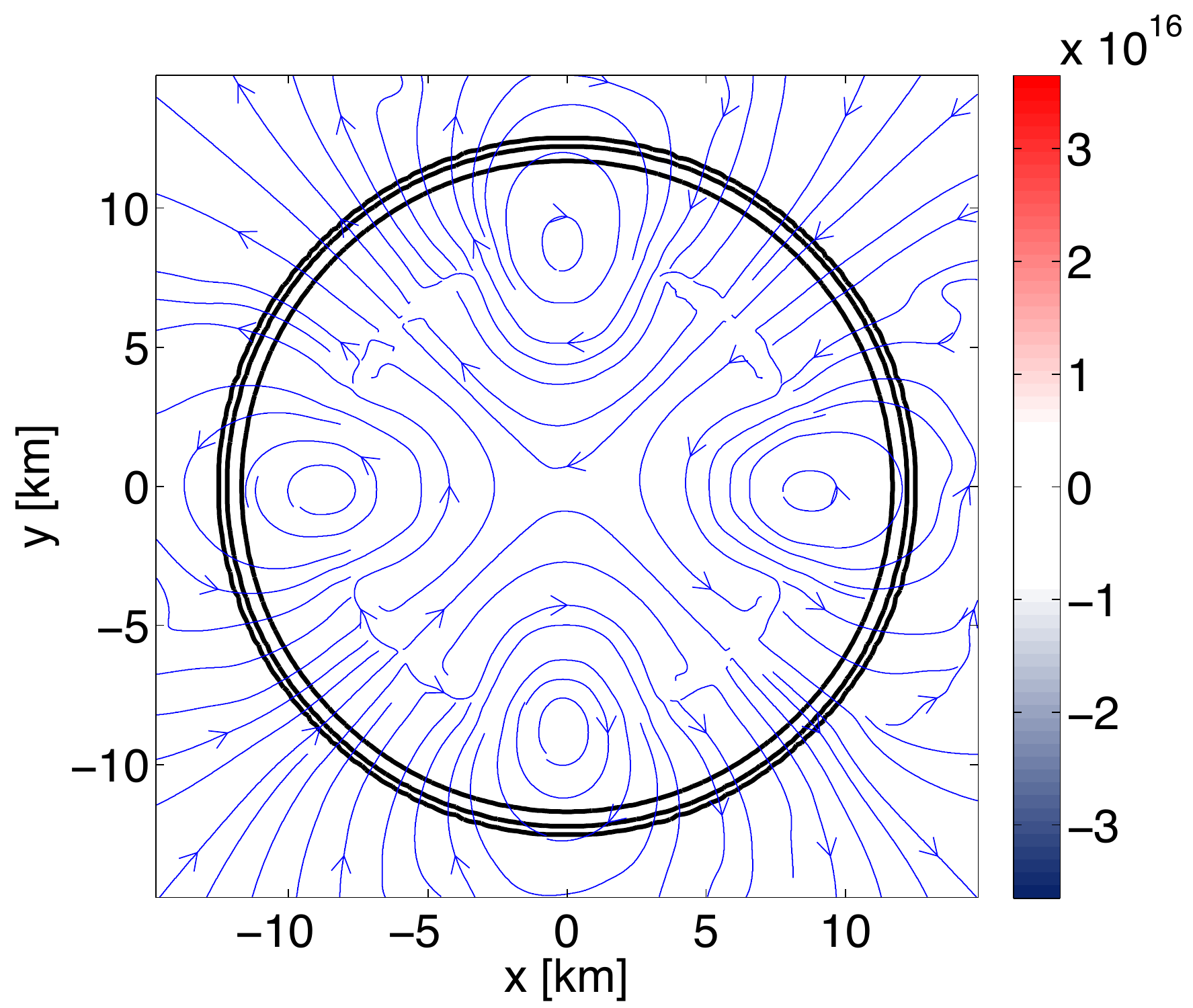}
     \hskip 0.cm
     \includegraphics[angle=0,width=5.4cm]{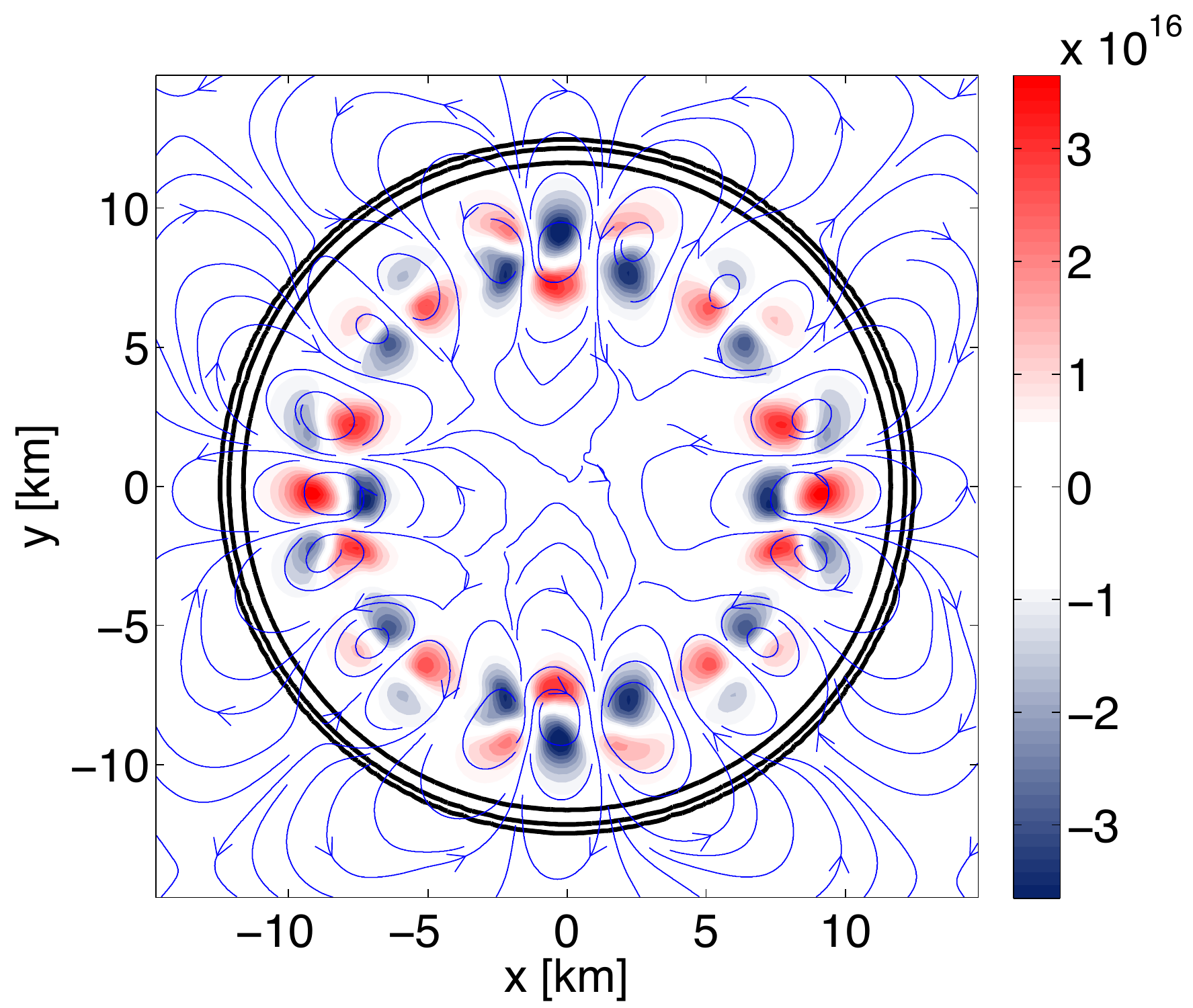}
     \hskip 0.cm
     \includegraphics[angle=0,width=5.4cm]{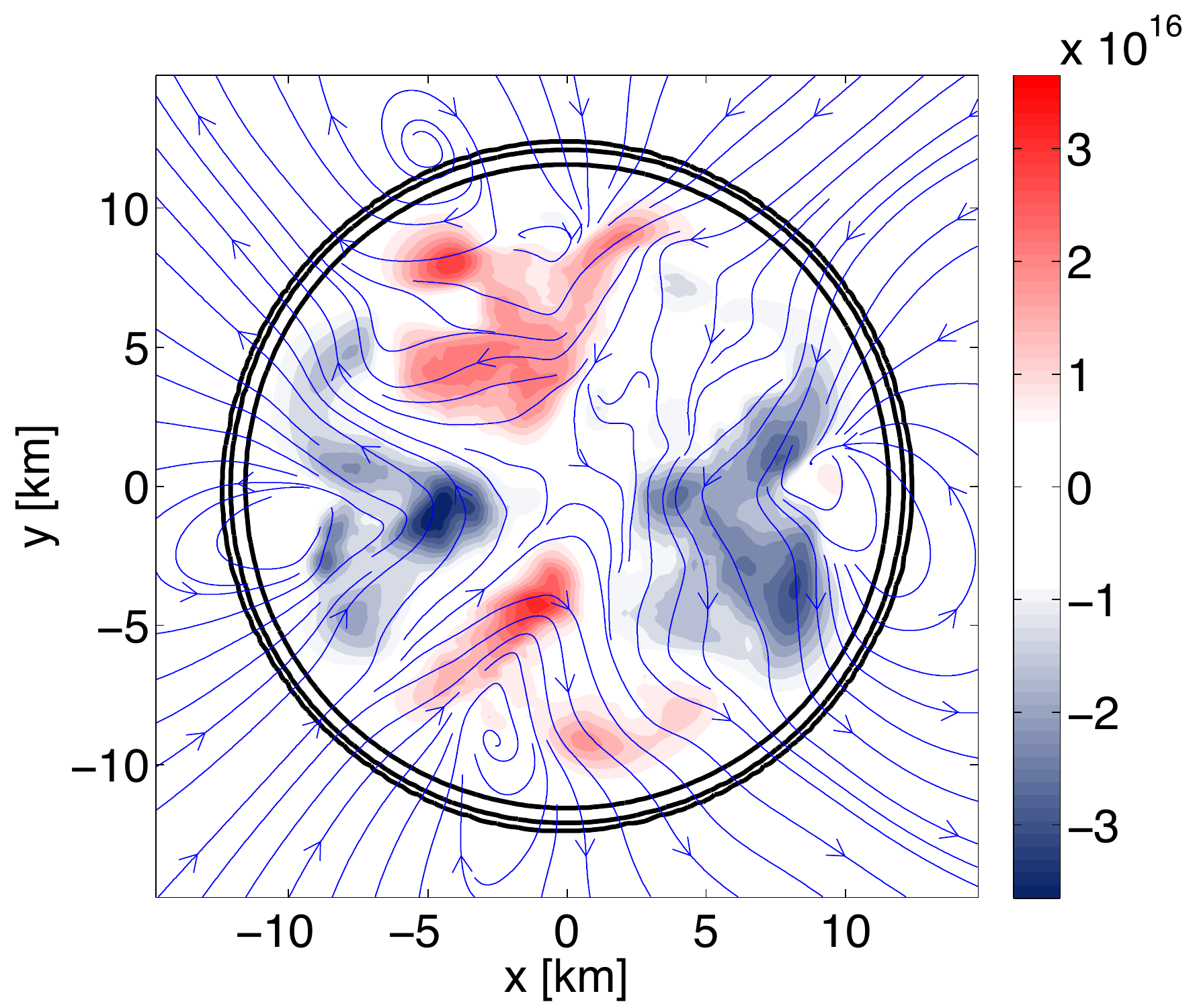}
  \end{center} \caption{Snapshots of the development of the
instability in our fiducial star, showing projections on the $(x,z)$
plane (upper row) and $(x,y)$ plane (lower row) of the simulation at
times $t=1, 3, 10\,{\rm ms}$ (left to right), respectively. Shown with
vector lines are the (global) magnetic-field lines, while the colours
show the intensity of the toroidal magnetic field only; also reported
are the iso-density contours of the rest-mass density near the stellar
surface.}
\label{fig:fig1}
\end{figure*}


\section{Physical system and numerical setup}

We model our initial NS as a nonrotating isolated fluid body
in ideal magnetohydrodynamics (MHD) and with a purely poloidal
magnetic field permeating it and extending to the exterior. The
initial axisymmetric equilibrium configuration is generated by the
\texttt{LORENE} code, which produces a fully relativistic solution and
consistently accounts for the metric distortions and structure
deformations induced by the magnetic field
(see~\citet{Bocquet1995}). The star is modelled as a polytrope with
equation of state $p\equiv K\rho^{\Gamma}$, where $\Gamma=2$ and
$K=98.5$, so that a NS with mass of $1.41\,M_\odot$ has a
radius of about $12.1\,{\rm km}$. Because the timescale for the
instability is shorter (and computationally feasible) for stronger
magnetic fields, we have considered stars with $B_{0}/10^{16}\,{\rm G}
\in [2.5, 10]$, where $B_{0}$ is the value at the magnetic pole,
selecting a value of $B_{0}=6.5\times 10^{16}\,$G as the reference
one.

Using these initial configurations, we perform general-relativistic
MHD simulations in three spatial dimensions under the Cowling
approximation, \ie we do not evolve the Einstein equations but
consistently solve for the MHD equations in a fixed and curved
spacetime. This choice is motivated by wanting to reduce computational
costs and by the fact that the changes in the spacetime are expected
to be intrinsically small. As a consequence, the GW emission is
computed using the Newtonian quadrupole
formula~\citep{Nagar2007,Baiotti:2008nf}. The evolutions are performed
with the \texttt{WhiskyMHD} code, whose properties have been tested
and discussed in a number of earlier papers~\citep{Giacomazzo:2007ti,
  Pollney:2007ss, Giacomazzo:2009mp, Giacomazzo:2010}. Our standard
numerical setup consists of a grid with three refinement
levels~\citep{Schnetter-etal-03b}, the highest one having a resolution
$h/M_{\odot}=0.17\simeq 250\,{\rm m}$ and covering all of the
star. The outer boundaries are placed at a distance of
$54\,M_{\odot}\simeq 79\,{\rm km}$.

The most salient difference of the code with respect to the references
above is in the treatment of the atmosphere. We recall that as
customary in relativistic hydrodynamics using finite-volume methods,
we surround the star with a low-density ``atmosphere'', whose dynamics
is prescribed by suitable boundary conditions. More specifically, the
rest-mass density there is set to a constant value, while the fluid
velocity is reset to zero (see~\citet{Baiotti04} for
details). Although this prescription works very well in hydrodynamic
simulations, it becomes problematic in ideal MHD, since it prevents
any evolution of the magnetic field in this region. For sufficiently
strong magnetic fields this approach can rapidly lead to errors at the
stellar surface, which prematurely terminate the simulations. To
improve on our treatment of the atmosphere and allow for a dynamics of
the magnetic field at surface and outside the star, we add a magnetic
diffusivity term to the induction equation, which we write as
$\partial_t (\tilde{B}^i) =
\partial_j(\tilde{v}^i\tilde{B}^j-\tilde{v}^j\tilde{B}^i) + \eta
\partial_i\partial^i \tilde{B}^j$, where $\eta$ is the scalar
resistivity (see~\citet{Giacomazzo:2007ti,Giacomazzo:2010} for details
on the implementation in ideal-MHD case). Because we want to retain
the ideal-MHD behaviour in the stellar interior and allow for an
evolution in the atmosphere, we set the resistivity to zero within the
bulk of the star, letting it increase continuously to its atmospheric
value starting from a low-density region near the stellar
surface. More specifically, we set $\eta(\rho) = \eta_0 f(\rho)$,
where $\rho$ is the rest-mass density, $f(\rho)$ is the Fermi function
and $\eta_0$ is a free parameter. We used a reference value of
$\eta_0/M_{\odot}=0.12$, but we will discuss how results change for a
lower value.

Again to reduce computational costs and because we are interested in
the development of magnetic-field instabilities, we add a perturbation
designed to trigger them. From the analysis of~\cite{Markey1973} we
expect a $\theta$-component of the velocity near the neutral line to
induce the fastest-growing instability. Hence, we introduce a
perturbation of azimuthal index $m=2$ in this quantity, such that the
relative change in the magnetic field is $10^{-3}$. We have verified
that the results do not change qualitatively for other choices of the
perturbation and that an instability develops even without a
perturbation.


\section{Results}

Before discussing the nonlinear development of the instability, it is
useful to recall the predictions of the perturbative studies
of~\cite{Markey1973} and~\cite{Wright1973} about its early growth. In
particular, we expect that: \textit{(i)} the instability should be
localised in the closed-field line region around the neutral line (\ie
where the poloidal magnetic field vanishes); \textit{(ii)} the
instability should occur after about an Alfv\'en timescale $\tau_A$
(if $\tau_A \sim 2R\sqrt{4\pi\langle\rho\rangle}/B_{0}$, where
$\langle\rho\rangle$ is the average rest-mass density and $R$ the
stellar radius, $\tau_A\sim 3$ ms for our fiducial model);
\textit{(iii)} the (exponential) growth rate of the instability should
scale linearly with magnetic field strength. As we will discuss, all
of these expectations are met.

Figure~\ref{fig:fig1} provides snapshots of the development of the
instability in our fiducial star, showing projections on the $(x,z)$
plane (upper row) and $(x,y)$ plane (lower row) of the simulation at
times $t=1, 3, 10\,{\rm ms}$ (left to right), respectively. These
correspond to early, mid and late stages of the evolution. Shown with
vectors lines are the (global) magnetic-field lines, while the colors
show the intensity of the toroidal magnetic field only; also reported
are the iso-density contours of the rest-mass density near the
stellar surface.

As expected, the instability develops around the neutral line (left
column), rapidly generating a toroidal magnetic field in this region
(middle column). The growth of this component continues until it
reaches a comparable strength to the poloidal one. At this point, the
growth proceeds much more slowly, and the magnetic field evolution is
less dramatic (right column), as the star evolves towards a new
equilibrium. As revealed by the different panels in
Fig.~\ref{fig:fig1}, the development of the instability breaks the
axisymmetry of the initial configuration, leading to a complex
structure with high azimuthal wave numbers (the $m=10$ component is
dominant in the middle column), which is eventually replaced by an
$m=2$ geometry at later stages (right column). While in this complex
evolution the (small) $m=4$ component is probably inherited from the
Cartesian coordinate system, it is interesting that the high-$m$ modes
develop despite the initial perturbation being an $m=2$ one. Note also
that the toroidal magnetic field produced by the instability is
concentrated in vortices (smaller at early times and larger at later
times) and that it changes sign both on meridian planes (see upper
row) and on the equatorial one (see lower row). This late-time
structure is different from the typical (axisymmetric) twisted-torus
discussed in previous works~\citep{Braithwaite2009, Ciolfi2009,
  Lander:2009}. We believe that this loss of axisymmetry follows from
the conservation of magnetic helicity in ideal
MHD~\citep{Woltjer1958}. Since purely poloidal (or purely toroidal)
magnetic fields have zero helicity and the latter has to be conserved
during a transformation in the ideal-MHD limit, the generation of a
vortex structure represents the natural way in which a newly-generated
toroidal field will not violate the initial zero-helicity of the
system.

A few additional remarks are worth making about
Fig.~\ref{fig:fig1}. The first one is about the evolution of the
magnetic field in the regions right outside the star, which is
essentially controlled by the resistivity there. Although the
reference value used, $\eta_0/M_{\odot}=0.12$, is rather high and
responsible for a considerable decay of the magnetic field, it also
allows for a smooth evolution and removes the development of the
discontinuities which would appear in the ideal-MHD limit. As we will
discuss later on, the qualitative behaviour of the instability is not
affected by the value of the resistivity or by the initial magnetic
field strength. The second remark is about the dynamics of the
toroidal magnetic field that, as it grows to become locally comparable 
with the poloidal one, it also moves towards the stellar surface,
where it can induce outflows of matter for smaller values of
$\eta_0$. While a more detailed discussion of this process will be
presented in a subsequent work, it is worth mentioning here that these
winds could eject considerable amounts of matter (\ie $\sim
10^{-5}-10^{-4}\,M_{\odot}$) and thus have direct connections with the
magnetar phenomenology.

\begin{figure}
  \begin{center}
     \includegraphics[angle=0,width=8.0cm]{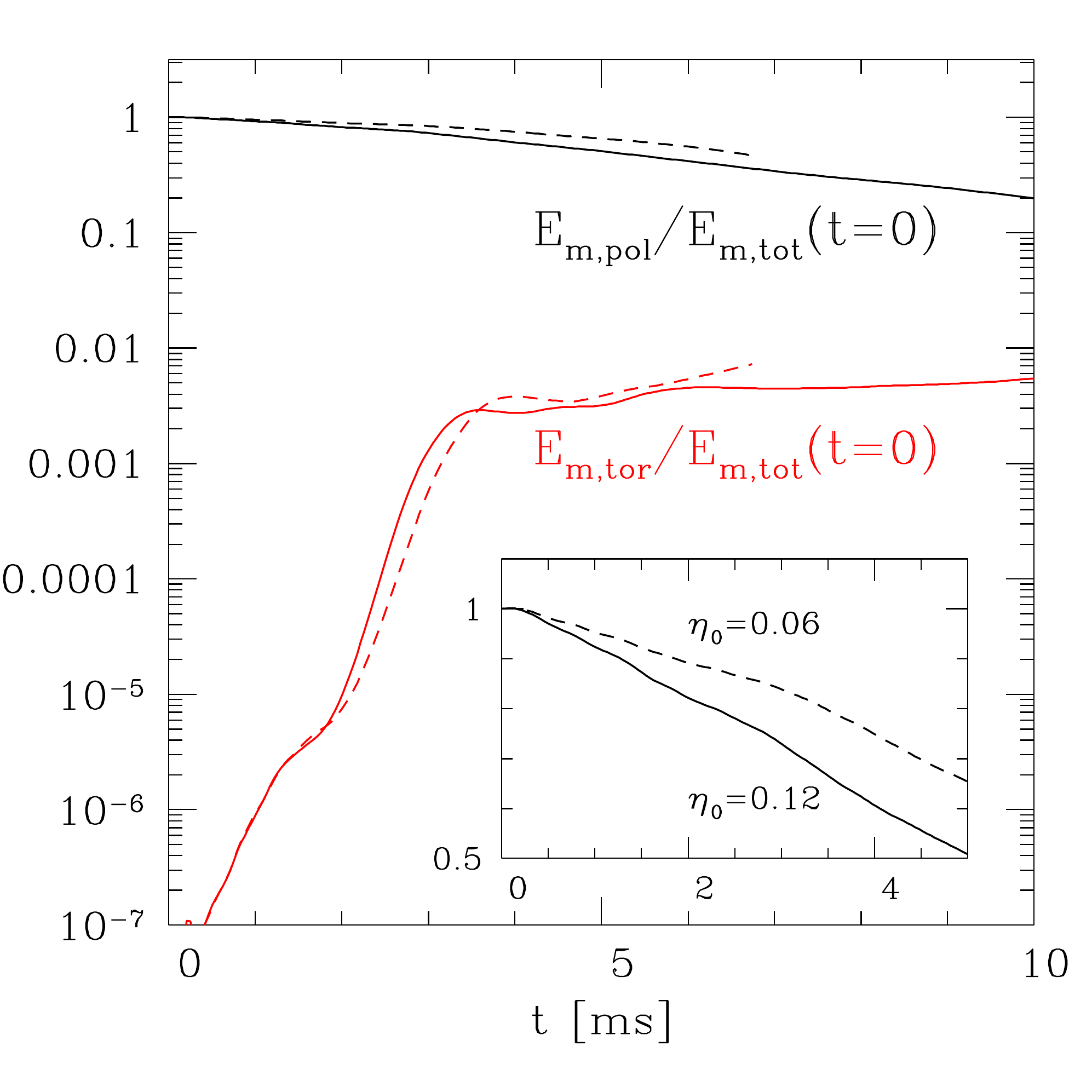}
  \end{center} \caption{Evolution of the energy in
the toroidal $E_{\mathrm{m,tor}}$ (red solid line) and poloidal
$E_{\mathrm{m,pol}}$ components (black solid line), normalised to the
initial value of the magnetic energy. The dashed lines refer to the
corresponding evolution with $\eta_0=0.06$ and show a smaller
dissipation of the magnetic field (see inset) but the same dynamics in
the instability.}
\label{fig:fig2}
\end{figure}

Figure~\ref{fig:fig2} provides a more detailed analysis of the
instability's behaviour, by monitoring the evolution of the energy in
the toroidal $E_{\mathrm{m,tor}}$ (red solid line) and poloidal
$E_{\mathrm{m,pol}}$ components (black solid line), normalised to the
initial value of the magnetic energy.  Initially the toroidal
component is very small, growing slowly until a time of $2$ ms. At
this point, which is close to our Alfv\'en-timescale estimate of $3$
ms, there is a sudden exponential growth, which lasts for another $1$
ms. The exponential growth, which nicely matches the predictions of
the linear-perturbation regime described by~\citet{Markey1973}, lasts
for $\sim 0.5\tau_A$ and then ceases, leaving a configuration which is
seen to be roughly unchanged for several more Alfv\'en
timescales. This suggests that the configuration shown in the right
column of Fig.~\ref{fig:fig1} is in a quasi equilibrium and is no
longer susceptible to the instability. Note that whilst the local
maxima of the field components are likely to dictate the system's
stability, the energy eventually present in the toroidal-field
component is only $\sim 3\%$ of the total magnetic energy. This value
may increase slightly on a much longer timescale.

Also reported in Fig.~\ref{fig:fig2} as dashed lines are the
corresponding evolution of the magnetic energies when a smaller
resistivity of $\eta_0/M_{\odot}=0.06$ is used. Since the evolution of
the instability in this case is qualitatively very similar (\cf the
evolution of $E_{\mathrm{m,tor}}$), we have confidence that our
prescription for the resistive behaviour of the magnetic field near
the stellar surface does not influence the dynamics of the
instability. At the same time, however, a smaller resistivity is also
responsible for a smaller decay of the poloidal magnetic field (see
inset), which is considerably dissipated by the end of the
simulation. While this behaviour is inevitable in a resistive context
and has been reported also by other authors~\citep{Braithwaite2007},
it represents an aspect of these evolutions which could be improved
with a fully consistent resistive MHD
approach~\citep{Palenzuela:2008sf}.

\begin{figure}
  \begin{center}
     \includegraphics[angle=0,width=8.0cm]{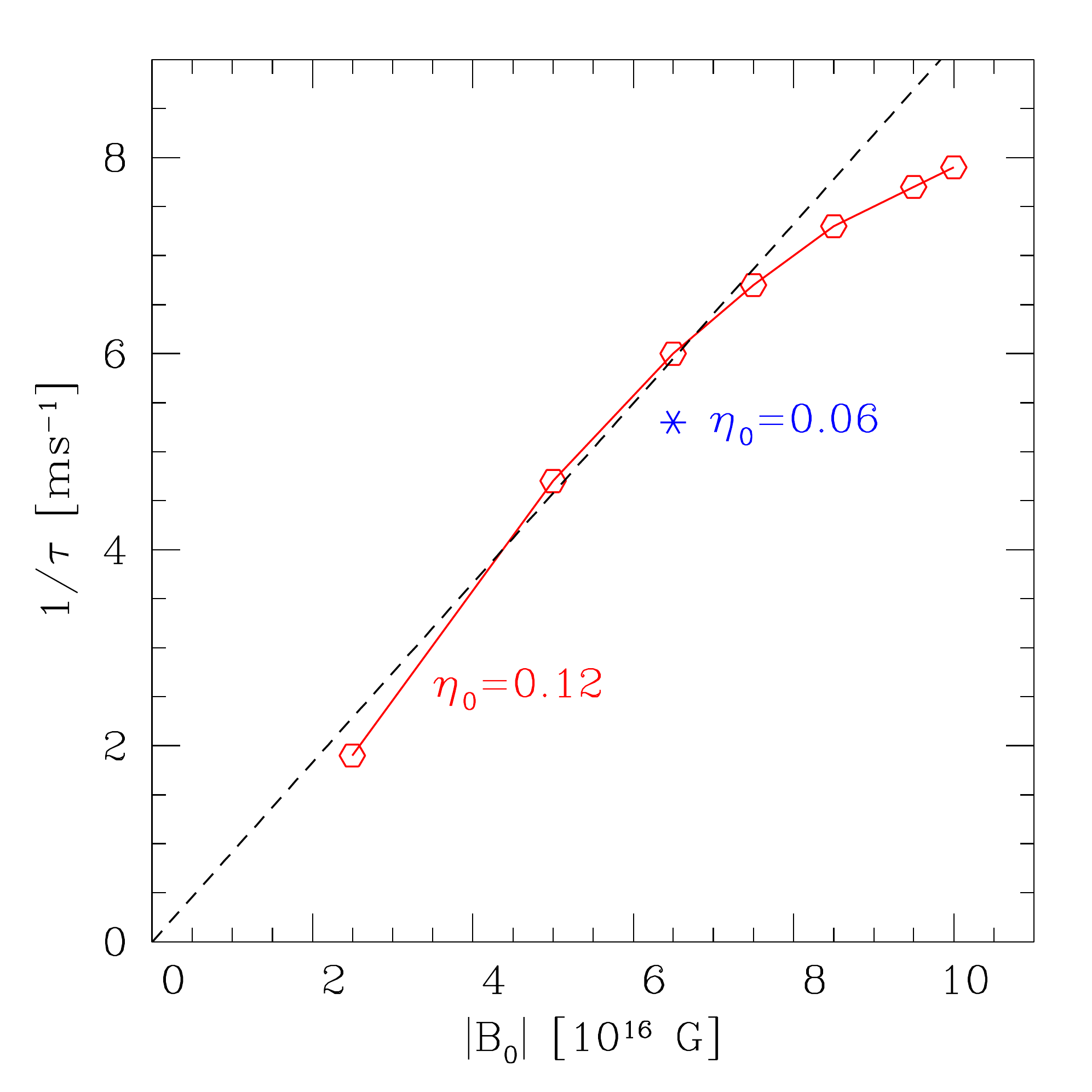}
  \end{center} \caption{Instability growth-rate as a
    function of the initial magnetic field strength for the NSs
    considered (red empty circles). Note the scaling remains
    approximately linear up to $B_0\simeq 7\times 10^{16}\,$G. Marked
    with a blue star is the corresponding growth-rate for $\eta_0=0.06$.}
\label{fig:fig3}
\end{figure}

Another important confirmation of the perturbative analysis is offered
in Fig.~\ref{fig:fig3}, where we show the inverse of the growth-time
$\tau$, defined through the exponential growth of the toroidal
component, versus the initial magnetic-field strength (red empty
circles). Note that the scaling is essentially linear for $B_0
\lesssim 7 \times 10^{16}\,$G, deviating from this for higher values,
because of the stronger magnetic tension. More specifically, the
stronger Lorentz force will tend to oppose the fluid motions in the
polar direction near the neutral line and which trigger the
instability. The presence of a linear scaling is essential to extend
our results to typical pulsar magnetic-field strengths, thus
estimating a growth-time of $\sim 10\,$s for a neutron star with
$B_0=10^{12}\,$G. Also marked in Fig.~\ref{fig:fig3} (blue star) is
the inverse growth-time for the fiducial star evolved with the smaller
resistivity of $\eta_0/M_{\odot}=0.06$; again, the close similarity in
the timescales confirms our expectation that the instability is not
influenced by the choice of the resistivity.

\begin{figure}
  \begin{center}
     \includegraphics[angle=0,width=8.0cm]{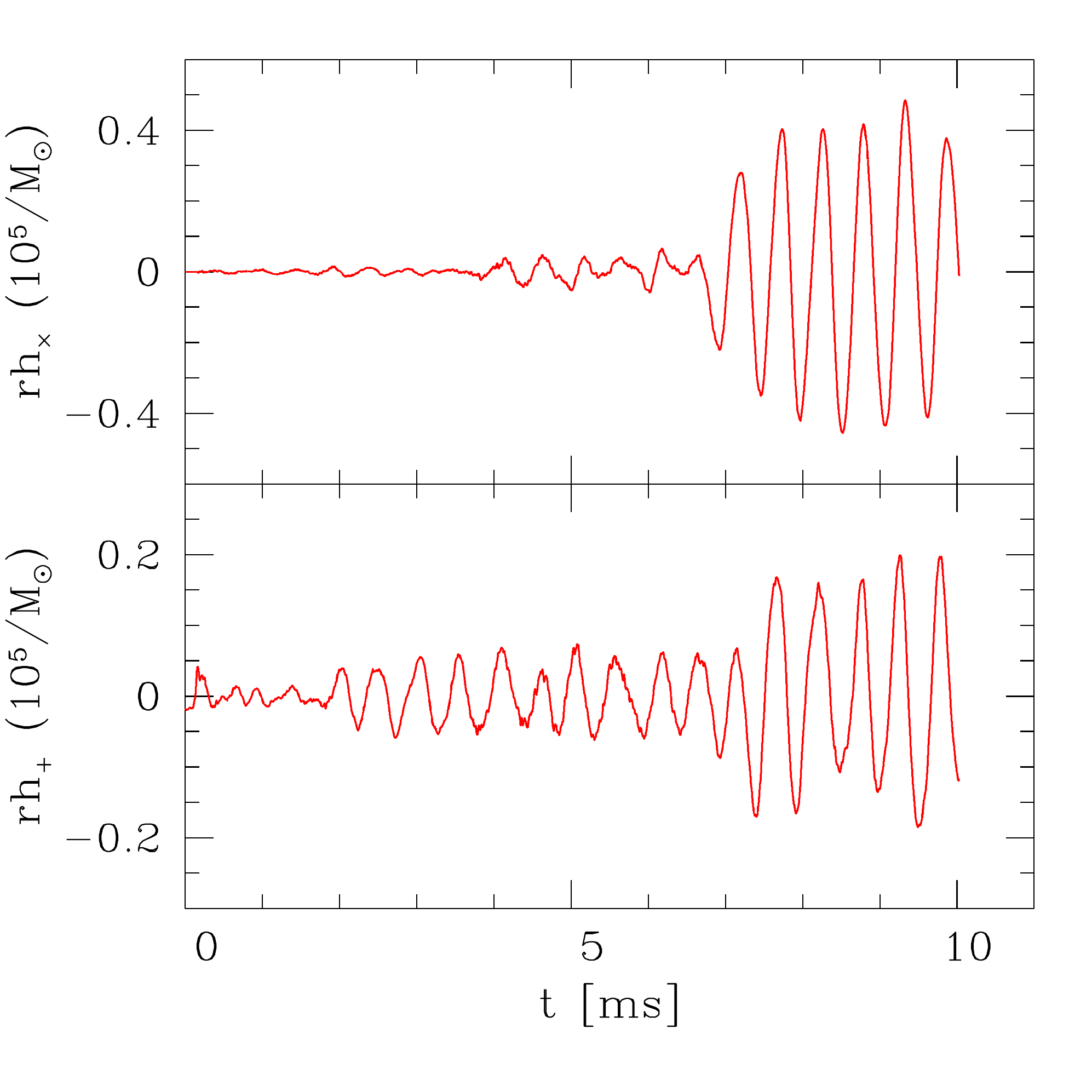}
  \end{center} \caption{GW strain in the $+$ and $\times$
    polarizations. Note the instability triggers large-amplitude
    $F$-mode oscillations.}
\label{fig:fig4}
\end{figure}

The final discussion is reserved for the potential GW signal emitted
during the development of the instability. In Fig.~\ref{fig:fig4} we
report the GW strain in the $+$ and $\times$ polarizations as computed
from the Newtonian quadrupole formula. It is quite apparent that the
signal is not of a burst type but, rather, that the main effect of the
instability is that of triggering large-amplitude oscillations of the
star in its fundamental $F$-mode. 

These GWs start emerging from the numerical noise already at $\sim
3.5\,$ms, but are associated to high-$m$ oscillations and hence not
efficient sources of GWs. However, as the magnetic field starts to
approach the final $m=2$ configuration at $\sim 7\,$ms, the
oscillations become more efficient in producing a GW signal (Note that
a $m=N$ pertubation in the magnetic field leads to a $m=2N$
perturbation in the density). Because these oscillations will have a
rather narrow spectral distribution peaked around the $F$-mode
frequency (which is not significantly affected by the presence of
magnetic fields), they represent very good sources of a periodic
signal, potentially detectable by future advanced detectors. Defining
the root-sum-square amplitude of the cross polarization as $h_{\rm
  rss}=\left[\int_{-\infty}^{+\infty} dt \,
  h_{\times}^2(t)\right]^{1/2}$, and assuming that the oscillations
will persist undamped for $\simeq 0.1-1\,$s, we estimate $h_{\rm rss}
= (0.54-1.7)\times 10^{-22}$ for a source at $10\,$kpc. The
corresponding signal-to-noise ratio for a detector such as
advanced-LIGO or advanced-Virgo is $S/N \simeq 1.6-5$, thus
potentially observable. A more detailed analysis of the spectral
properties of the GW signal will be presented in a future work.
These waveforms represent the first estimate of the conversion of the
kinetic energy generated through the instability into GWs. For weaker
magnetic fields, perturbative analyses have suggested this coupling is
much weaker~\citep{Levin:2011}, but more work is needed to investigate
nonlinearly this regime.

\section{Summary}

We report on numerical evolutions of the instability of poloidal
magnetic fields in relativistic stars and the subsequent generation of
a mixed-field configuration in quasi-equilibrium. In agreement with
the expectations from analytic perturbative studies~\citep{Markey1973,
  Wright1973}, we show that the instability appears after about an
Alfv\'en timescale, localised in the region of closed field lines. In
addition, the growth-rate of the instability has the expected linear
scaling with magnetic field up to very large field strengths.

The instability leads to the generation of a toroidal magnetic field,
which starts with a complex and high-$m$ azimuthal structure, and is
produced in local vortices. The instability ceases when the two field
components are locally comparable in strength and the resulting
configuration relaxes towards a geometry dominated by an $m=2$
toroidal component.  This non-axisymmetric configuration is different
from the simpler axisymmetric ones of earlier
studies~\citep{Ciolfi2009, Lander:2009, Braithwaite2009,
  Ciolfi2010}. However, it also represents the natural evolution of a
system which is required to conserve its initial zero helicity in the
ideal-MHD limit.

The total magnetic energy of the final configuration has around a
$3\%$ contribution from the toroidal component; this is comparable
with the equilibria studied by~\citet{Ciolfi2009}
and~\citet{Lander:2009}, but considerably lower than that
of~\cite{Braithwaite2009}, where stratified main-sequence stars were
considered. Interestingly, the development of the instability also
leads to the buoyancy of the newly-generated toroidal magnetic field
and this can result into a mass outflow near the stellar
surface. Additional work is needed to establish how these winds can be
related to the phenomenology observed in giant flares of magnetars.

A consequence of the instability for the very strong magnetic fields
considered here is that of triggering large-amplitude oscillations of
the star in its $F$-mode. Our simulations show that the
small-amplitude oscillations of the star are amplified by about an
order of magnitude by the time the instability has saturated. Because
these oscillations will have frequencies around the $F$-mode
frequency, it is reasonable to perform searches for periodic signals
associated to giant flares in magnetars. For a source at $10\,$kpc
with oscillations persisting undamped for $\simeq 0.1-1\,$s, the
root-sum-square amplitude at $1500\,\hz$ will be $h_{\rm rss} =
(0.54-1.7)\times 10^{-22}$, thus leading to a signal-to-noise ratio
$S/N \simeq 1.6-5$ for a detector such as advanced-LIGO or
advanced-Virgo.

\medskip
During the completion of this work we have become aware of a very
similar analysis carried out by~\citet{Lasky2011}, where the
instability of purely poloidal magnetic fields in relativistic stars
was also presented. Despite the different numerical setups, the
results obtained by~\citet{Lasky2011} about the development of the
Tayler instability are in good agreement with those presented here,
thus validating each other's conclusions.

\bigskip
We are grateful to Bruno Giacomazzo, Ian Jones and Yuri Levin for
useful comments, and Roberto De Pietri for help in the GW
estimates. Support comes also from the ``Della Riccia'' Foundation,
from ``CompStar'', a Research Networking Programme of the European
Science Foundation, and from the DFG grant SFB/Transregio~7.



\end{document}